\documentclass{emulateapj}
\usepackage{multirow}

\def \vv#1{\mathbf{#1}}
\def \be {\begin{equation}}
\def \ee {\end{equation}}

\def \Da {D}
\def \Dc {D_\mathrm{crit}}

\def\crm{\cr\noalign{\medskip}}

\def \llabel#1{\label{#1}}

\def\aj{Astron.\ J. }
\def\apj{Astrophys.\ J. } 
\def\apjl{Astrophys.\ J. }

\def\aap{Astron.\ Astrophys. }

\def\epsl{Earth Planet.\ Sci.\ Lett. }

\def\grl{Geophys.\ Res.\ Lett. }

\def\jgr{J.\ Geophys.\ Res. }

\def\nat{Nature }

\shorttitle{Impact cratering on Mercury}
\shortauthors{Correia \& Laskar}

\begin{document}


\title{Impact cratering on Mercury: consequences for the spin evolution} 
%\title{Impactor flux and cratering on Mercury: \\ 
%consequences for the spin evolution} % and early Solar System}


\author{Alexandre C.M. Correia}
\affil{Department of Physics, I3N, University of Aveiro, Campus de
Santiago, 3810-193 Aveiro, Portugal;}
%\affil{ASD, IMCCE-CNRS UMR8028,
%Observatoire de Paris, UPMC,
%77 Av. Denfert-Rochereau, 75014 Paris, France}
%\email{correia@ua.pt} 

\and

\author{Jacques Laskar}
\affil{ASD, IMCCE-CNRS UMR8028,
Observatoire de Paris, UPMC,
77 Av. Denfert-Rochereau, 75014 Paris, France}


\begin{abstract}
Impact basins identified by Mariner 10 and Messenger flyby images provide us a fossilized record of the impactor flux of asteroids on Mercury during the last stages of the early Solar System.
The distribution of these basins is not uniform across the surface, and is consistent with a primordial synchronous rotation \citep{Wieczorek_etal_2012}.
By analyzing the size of the impacts, %we show that the distribution for asteroid diameters $ D < 110 $~km is compatible with an index power law of $-1.2$. %, a value that matches the predicted primordial distribution of the main-belt.
we derive a simple collisional model coherent with the observations. When
combining it with the secular evolution of the spin of Mercury, we are able to reproduce the present 3/2 spin-orbit resonance ($\sim$50\% of chances), as well as a primordial synchronous rotation.
%This result is independent of dissipation models, and valid for both prograde and retrograde initial rotation.
This result is robust with respect to variations in the dissipation and collisional models, or in the initial spin state of the planet.
%Our result is very robust and model independent. %, which sets important constraints for the initial size distribution of asteroids in the main-belt.
\end{abstract}


%% subject header.  Objects should be in the appropriate "individual"
%% headers (e.g. quasars: individual, stars: individual, etc.) with the
%% additional provision that the total number of headers, including each
%% individual object, not exceed six.  The \objectname{} macro, and its
%% alias \object{}, is used to mark each object.  The macro takes the object
%% name as its primary argument.  This name will appear in the paper
%% and serve as the link's anchor in the electronic edition if the name
%% is recognized by the data centers.  The macro also takes an optional
%% argument in parentheses in cases where the data center identification
%% differs from what is to be printed in the paper.

\keywords{
minor planets, asteroids: general ---
planets and satellites: individual (Mercury) ---
planets and satellites: dynamical evolution and stability}
% --- methods: analytical, numerical}


\section{Introduction}

Mercury is known to be in a  3/2 spin-orbit resonance \citep{Pettengill_Dyce_1965, Colombo_1965,Goldreich_Peale_1966}.
However, recent investigation on the locations of ancient impact basins on Mercury (pre-Caloris basin) has shown that they are not uniformly distributed across the surface, and
that their distribution is consistent with the spatial variations that would arise if Mercury were once in a state of synchronous rotation \citep{Wieczorek_etal_2012}.
Assuming an initial prograde rotation for Mercury, the 3/2 resonance arises naturally, but the synchronous rotations is almost impossible to achieve ($< 4$\% of chances) due to the high values of Mercury's orbital eccentricity \citep{Correia_Laskar_2004,Correia_Laskar_2009}.
An alternate scenario for capture into the 1/1 resonance is to suppose that Mercury's initial spin was retrograde, since planetary accretion models seem to allow  terrestrial planets to have either prograde or retrograde initial spin rates  \citep{Dones_Tremaine_1993, Kokubo_Ida_2007}.
In this case, synchronous rotation  becomes the most likely outcome, but a subsequent evolution into the presently observed 3/2 resonant rotation requires a large impact event  \citep{Wieczorek_etal_2012}.

In the present  Letter, we extend further the dynamical analysis of asteroid  impacts on the planet surface.  We obtain the impactor flux on Mercury from the present crater distribution (Sect.\,2), and provide two simple models for collisions (Sect.\,3).
Then, using a secular spin evolution model for Mercury (Sect.\,4) with collisions, we determine the chances of capture in each spin-orbit resonance (Sect.\,5), and derive our conclusions (Sect.\,6).

\section{Impact cratering}

Cratering is one of the most important geological processes that shape and modify the surfaces of terrestrial planets and satellites.
The quantification of the impactor source population on Mercury and the observation of its surface help to understand the dynamical history of both the impactor population and the spin of the planet.

Impact basins identified by Mariner 10 based geological mapping \citep{Spudis_Guest_1988} and Messenger flyby images \citep{Fassett_etal_2011, Herrick_etal_2011} provide us a fossilized record of the impactor flux of asteroids during the last stages of the early Solar System. 
%The majority of these basins are older than the Caloris basin, the largest among all, whose age is estimated at 3.73~Gyr \citep{LeFeuvre_Wieczorek_2011}.
The spatial density of craters smaller than 100~km in diameter have been affected by more recent geologic processes and the numerous craters that are somewhat larger than this might also be in saturation \citep{Fassett_etal_2011}. 
In addition, the minimal crater size to escape the presently observed 3/2 spin-orbit resonance is estimated to be 300~km \citep{Wieczorek_etal_2012}.
Therefore, here we restrict our analysis to those basins that are larger than 300~km (Table\,\ref{Tab1}).

\begin{table}
\begin{center}
\caption{Impact basins on Mercury with diameters greater than 300~km \citep{Wieczorek_etal_2012}. \llabel{Tab1}}
\begin{tabular}{l r r r r}
\tableline
\tableline
Crater &	Lat. & Long. & $D_\mathrm{basin} $ & $\Da$ \\
 (name)  & ($^\circ$N) & ($^\circ$E) & (km) & (km)  \\
\tableline
Caloris &	$+30.9$ & $+159.7$ &1456.7 & 188.9 \\
Andal-Coleridge & $-43.0$ & $-49.0$ & 1300.0 & 163.8 \\
Tir &	$+6.0$ & $-168.0$ & 1250.0 & 156.0 \\
Eitoku-Milton &	$-23.0$ &	$-171.0$	& 1180.0	& 145.2 \\
Bartok-Ives & $-33.0$ & $-115.0$ &	1175.0 & 144.4 \\
Donne-Moliere & $+4.0$ & $-10.0	$ & 1060.0 & 126.9 \\
Sadi-Scopas &	$-83.0$ & $-44,0 $ & 930.0 & 107.8 \\
Budh & $-17.0 $ & $-151.0$ & 850.0 & 96.3 \\
Matisse-Repin & $-23.4$ & $-75.2$ & 843.2 & 95.4 \\
Mena-Theophanes & $-2.3$ & $-126.7$ & 836.1 & 94.4 \\
Sobkou & $+33.4$ & $-133.5$	& 785.3 & 87.3 \\
Borealis & $+72.1 $ & $-80.9 $ & 785.2 & 87.2 \\
Rembrandt & $-33.1$ & $+87.7$ & 696.7	& 75.1 \\
Vincente-Yakovlev & $-52.6 $ & $-162.1$  &692.5 & 74.6 \\
Ibsen-Petrarch & $-31.0 $ & $-30.0$ & 640.0 & 67.6 \\
Beethoven & $-20.8$ & $-123.9$ & 632.5 & 66.6 \\
Brahams-Zola & $+59.0 $ & $-172.0$ & 620.0 & 64.9 \\
(unnamed) & $+4.7$ & $+74.1$ & 529.6 & 53.3 \\
Tolstoj & $-17.1 $ & $-164.6$ & 500.6 & 49.7 \\
Hawthorne-Riemen. & $-56.0$ & $-105.0$ & 500.0 & 49.6 \\
Gluck-Holbein & $+35.0$&$-19.0$ & 500.0 & 49.6 \\
(unnamed) & $+0.6 $ & $+93.4$  & 428.4	& 40.9 \\
(unnamed) & $-39.0	$ & $-101.4$ & 420.3 & 39.9 \\
Dostoevskij & $-44.8$ & $-177.1$ & 413.9 & 39.2 \\
(unnamed) & $-44.5	$ & $-93.2$ & 411.4 & 38.9 \\
Derzhavin-Sor Juana & $+50.8$ & $-26.9$ & 406.3 & 38.3 \\
(unnamed) & $-2.6 $ & $-56.1 $ & 392.6 & 36.7 \\
(unnamed) & $+27.9$ & $-158.6 $ & 389.0 & 36.3 \\
Vyasa & $+50.7$ & $-85.1$ & 379.9 & 35.2 \\
Shakespeare & $+48.9 $ & $-152.3$ & 357.2 & 32.6 \\
Hiroshige-Mahler & $-17.0$ & $-23.0$ & 340.3 & 30.7 \\
Chong-Gauguin & $+57.1$  & $-107.9$ & 325.6 & 29.0 \\
Raphael & $-20.3$ & $-76.1$ & 320.4 & 28.5 \\
Goethe & $+81.5$ & $-54.3$ & 319.0 & 28.3 \\
(unnamed) & $-2.5$ & $-44.6$ &  311.4 & 27.5 \\
(unnamed) & $+28,9$ & $	 -113.8$ & 307.9 & 27.1 \\
(unnamed) & $-25.0$ & $-98.8 $ & 307.6 & 27.0 \\
(unnamed) & $-17.3$ & $-96.8$ & $303.4$ & 26.6 \\ \tableline
\end{tabular}
%\caption{Data taken form http://exoplanets.org/}
\tablecomments{The diameter of the impactor  is given by Eq.(\ref{120304b}). }
\end{center}
\end{table}

Cratering processes have been extensively studied through impact and explosion experiments \citep[e.g.][]{Schmidt_Housen_1987}. By interpreting the observations of the Deep Impact event,
\citet{Holsapple_Housen_2007} derived an expression that allow us to estimate the diameter of a basin, $D_\mathrm{basin}$, as a function of the diameter of the impactor, $\Da$.
For non porous rocks, we have \citep{LeFeuvre_Wieczorek_2011}:
\be
D_\mathrm{basin} \approx 1.92 \, \Da \left( \frac{u^2} {g \Da} \right)^{0.22} \left( \frac{\rho_a}{\rho} \right)^{0.31}   \ , \llabel{120304a}
\ee
where $\rho$ and $\rho_a$ are respectively the densities of the planet and the impactor, $g$ is the surface gravity of the planet, and $u$ the normal impact velocity component.
Adopting Mercury's values for $\rho$ and $g$, a mean density for the asteroids $\rho_a = 2.5 $~g/cm$^3$ \citep{Fienga_etal_2011}, and an average speed of $u = 42.5$~km/s \citep{LeFeuvre_Wieczorek_2008}, we can estimate the size of the bodies that  opened the basins currently observed (Table\,\ref{Tab1}):
\be
{\Da}_\mathrm{[km]} \approx \left( \frac{{D_\mathrm{basin}}_\mathrm{[km]}}{22} %{22\,\mathrm{km^{0.2}}} 
\right)^{1.25}  \ . \llabel{120304b}
\ee

We can now derive the cumulative distribution of the impact number on Mercury as a function of the size of the bodies. 
In Figure\,\ref{F1} we used bins of 5~km and each dot gives the position of the bin.
This distribution provides a reliable estimation of the impactor flux during the late heavy bombardment some 3.8~Gyr ago \citep{Kring_Cohen_2002}.
We simultaneously plot the present size distribution in the asteroid-belt \citep[e.g.][]{Jedicke_etal_2002}, normalized by a factor 50, such that the number of largest bodies coincide with the observed number of impacts on Mercury.
Although the two populations distributions are quite similar for $ \Dc > 110 $~km, for smaller diameters it appears that the number of impacts is clearly smaller.

\begin{figure}%[htb]	% h-here, t-top, b-bottom
\centering
\includegraphics[width=8.cm]{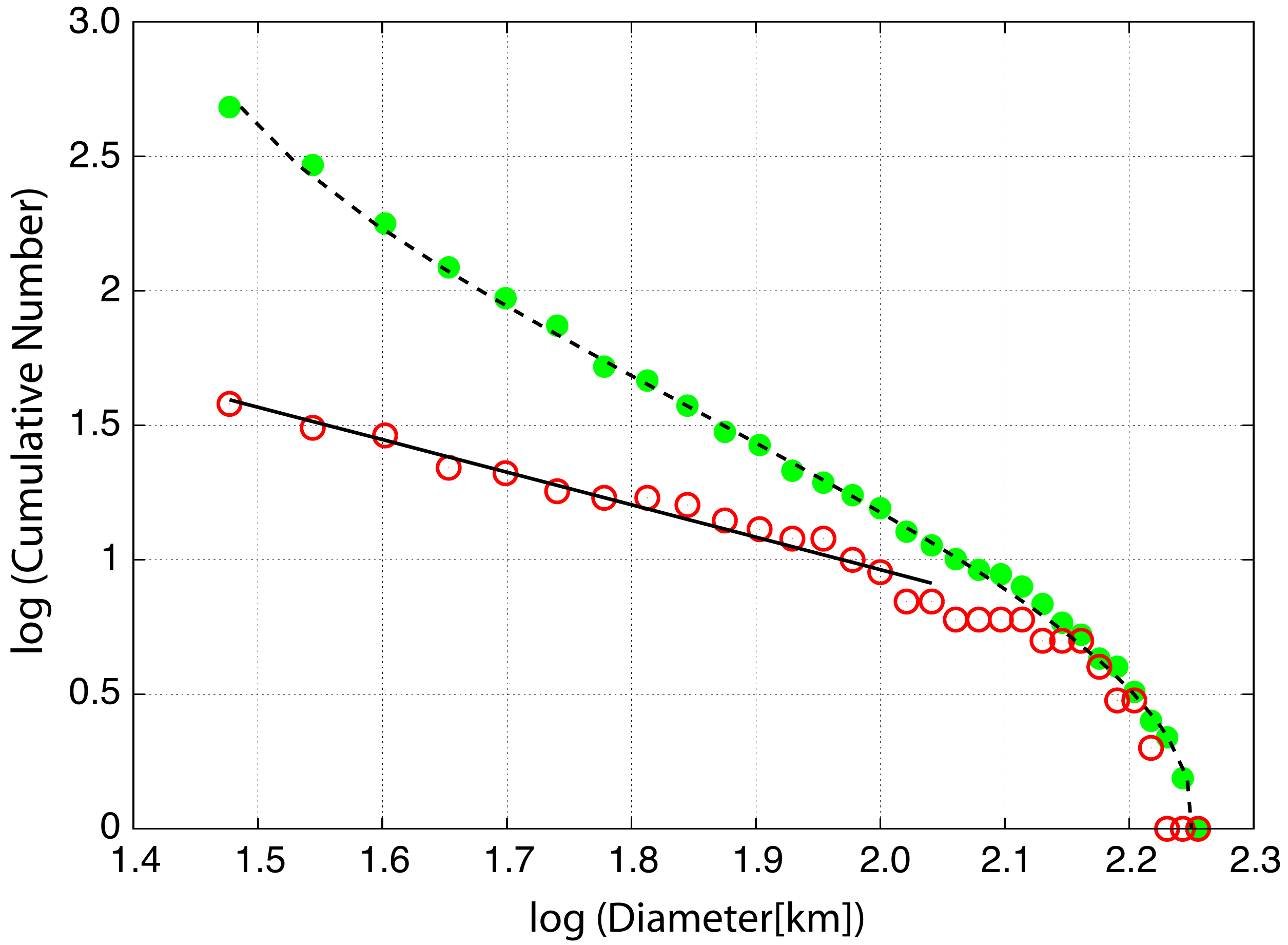} %	** if .eps don't need extension
\caption{Cumulative number of impacts observed on Mercury's surface (IMD, red), and normalized size distribution in the asteroid-belt (ABD, green). The diameter of the bodies is binned in classes of 5~km.
%For diameters $ D \le 110 $~km the impact population follows an incremental power-law index of $-1.2$ (solid line).
\llabel{F1}}
\end{figure}

%Indeed, for larger diameters it is believed that breakup events among asteroids do not occur frequently enough to significantly modify the shape of the population \citep{Bottke_etal_2005a}.
%On the other hand, for smaller diameters it is expectable that the impact distribution on Mercury gives an indication of the primordial main-belt distribution.
%Interestingly, both values exactly match the theoretical initial size distribution for the main-belt given by \citet{Bottke_etal_2005a}. 
%Although the index $-4.5$ is subject to large incertitude and it is sensitive to changes in $\Dc$, the lower index $-1.2$ is very robust and provides a reliable information on the primordial main-belt distribution.
%Similar studies on Ceres and Vesta should confirm these observations \citep{deElia_DiSisto_2011}.

\section{Collisional models}
\llabel{sectcoll}

A realistic long-term evolution of the spin of Mercury can only be complete
if one takes into account the effect from collisions during the early stages of its evolution.
Escape from spin-orbit resonances can occur by variations in the eccentricity \citep{Correia_Laskar_2004,Correia_Laskar_2009}, but evolution beyond these states is also possible through the momentum imparted to the planet during a basin-forming impact event \citep[e.g.][]{Melosh_1975}. 
In particular, the synchronous resonance can only be unlocked by this method \citep{Lissauer_1985,Wieczorek_LeFeuvre_2009}.

The presently observed craters on the planet's surface provide us important
information on the size and distribution of the impacts, but few on their
consequences to the spin of the planet. % at the time. 
Therefore, we adopt here a simple model to take into account large impacts on
Mercury, that uses either the observed crater or asteroid distributions to generate similar 
random impacts. 

%First, 
We assume that the impact probability is uniformly distributed over the whole surface of the
planet.
The rotational angular momentum of Mercury prior to the collision is given
by $ \vv{L}_0 = C \omega_0 \vv{k}_0 $, where $C$ is the principal moment of inertia, $\omega $ the rotation rate, and $ \vv{k} $ a unit-vector along the spin axis.
The impactor has mass $m_i$ and hits
the planet at a point of the surface given by $ \vv{R} $.% , where $ R = || \vv{R} || $ is the radius of the planet. 
The planet has mass $m$, and assuming $ m_i \ll m $ the angular momentum change is simply obtained by
\be
\vv{L} = C \omega \vv{k} = C \omega_0 \vv{k}_0 + m_i \vv{R} \times \vv{u} \ , \llabel{111024a}
\ee
where $ \vv{u} $ is the differential velocity between Mercury and the impactor. %(Fig.\,\ref{figimpactor}).

%Second, 
We assume also for simplicity that the orbit of Mercury is circular $(r = a)$ and
coplanar with the orbit of the impactor.
Whenever the position of the impactor $ \vv{r}_i $ crosses the orbit of Mercury,
a collision can occur, that is, when $ r_i = a $.
Thus, we compute the relative velocity as
\begin{eqnarray}
u = n a \sqrt{3 - \frac{a}{a_i} -2 \sqrt{\frac{a_i}{a} (1-e_i^2)}} 
 \ , \llabel{111024c}
\end{eqnarray}
where $a$ is the semimajor axis, $e$ the eccentricity, and $n$ the mean motion.
Based on the present distribution of Mercury-crossing objects\footnote{\it http://www.minorplanetcenter.net/}, % http://www.minorplanetcenter.net/iau/MPCORB/MPCORB.DAT
we assume that the impactor has origin in the asteroid belt, thus
$ a_i = a_A / (1+e_i) $, and $ e_i \ge (a_A-a)/(a_A+a) $, where $ a_A \approx 2.5 $~AU.
% for the asteroid belt. %is the semi-major axis of the asteroid belt.

Finally, we assume that the initial obliquity of the planet is zero, i.e., $\vv{k}_0 $ is normal to the orbital plane. %In order to describe the collisions, we use 
Using for simplicity the reference frame $ ( \vv{i}_0,\vv{j}_0,\vv{k}_0 ) $ linked to the orbital plane of Mercury, where $ \vv{j}_0 $ is along the direction of $\vv{u}$, we have
\be
\vv{R} \times \vv{u} = - R_a u \sin \phi \, \vv{i}_0 + R_a u
\cos \phi \, \vv{k}_0 \ , \llabel{111024e}
\ee
where $ R_a = || \vv{R} \times \vv{u} || / u $ is the moment arm, which is
perpendicular to $ \vv{u} $, and $ \phi $ is the angle between $ \vv{i}_0 $  and the impact point direction. %(Fig.\,\ref{figimpactor}).
Using expression (\ref{111024a}), the change in the rotation rate after impact is %estimated as:
\be
\left(\frac{\omega}{\omega_0}\right)^2 = 1 + \ell^2 + 2 \ell  \cos \phi \ ,
\llabel{111024f}
\ee
where
$ \ell = m_i R_a u / (C \omega_0) $.
In Table\,\ref{Tab2} we list the minimal impactor diameter needed to disrupt each
spin-orbit resonance for different eccentricities ($R_a=R$ and $\phi=0^\circ$) with an average speed of $ u = 42.5 $~km/s \citep{LeFeuvre_Wieczorek_2008}.

\begin{table}
\begin{center}
\caption{Critical eccentricity and minimal impactor diameter needed to destabilize each spin-orbit resonance. % for the present eccentricity. 
\llabel{Tab2}}
\begin{tabular}{c c | c c c}
\tableline
\tableline
\multirow{2}{*}{$p$} & \multirow{2}{*}{$ e_\mathrm{crit.} $}  &  \multicolumn{3}{c}{$\Da$ (km)} \\
 & & $e=0.1$ & $e=0.2$ & $e=0.3$  \\
\tableline
   5/1 &      0.211334 &$-$ & $-$ & 20.4 \\
   9/2 &      0.174269 &$-$ & 14.9 & 23.0 \\
   4/1 &      0.135506 &$-$ & 17.9 & 26.2 \\
   7/2 &      0.095959 & 12.2 & 21.4 & 29.4 \\
   3/1 &      0.057675 & 16.3 &25.5 & 32.7 \\
   5/2 &      0.024877 & 21.5 &30.0 & 36.0\\
   2/1 &      0.004602 & 27.9 & 34.8 & 39.1 \\
   3/2 &      0.000026 & 35.4 & 39.3 & 41.2 \\
   1/1 &       $-$           & 40.6 & 41.6 & 42.1 \\ 
   1/2 &      0.000180 & 25.7 & 28.8 & 30.8 \\ \tableline
\end{tabular}
\end{center}
\end{table}

The quantities $(m_i, e_i)$ depend on the impactor and $(R_a, \phi)$
on the impact point.
They are the only variables unknown in the model and they need to be randomized
in order to simulate the collisions.

We assumed that impacts over the surface of Mercury are uniformly distributed. 
Thus, the probability density function (PDF) for $\phi$ is constant in the interval $[
0, 2 \pi ]$, while for the moment arm $R_a$ the PDF is linear in the interval $[0, R]$
\citep{Laskar_etal_2011}. 
If $x$ is a random variable uniform in the interval $[0, 1]$ we then compute:
\be
\phi = 2 \pi x \ , \quad R_a = R \sqrt{x} \ . \llabel{111025a} 
\ee

For the periastron of the impactor $p_i = a_A (1-e_i) / (1+e_i)$,
we use a similar distribution as for $R_a$, that is, we assume that the PDF for $p_i$ is
linear in the interval $[0, a]$:
\be
p_i = a \sqrt{x}  \quad \Rightarrow \quad e_i = \frac{a_A - a \sqrt{x}}{a_A + a
\sqrt{x}} \ . \llabel{111025b}
\ee

Finally, for the mass distribution we use either the present asteroid-belt distribution (ABD), or the observed impacts on Mercury distribution (IMD).
The total number of collisions $N$ is a parameter that can be adjusted in the model,
%but the probabilities are normalized in order to ensure that they always match the distributions shown in Figure\,\ref{F1}.
and the corresponding distribution of the masses is given by $ m_i = \pi \rho_a
\Da^3 / 6 $.

The ABD can be fitted by a polynomial curve of degree three (dashed line, Fig.\,\ref{F1}):
\be
\log \Da = 0.06 \, y^3 - 0.28 \, y^2 + 0.03 \, y + 2.25
\ , \llabel{120408a}
\ee 
where $ y = \log (x N) $.

The IMD appears to have a bump near $ \Dc \sim 110$~km, that can be explained by different collisional regimes in the asteroid-belt \citep[e.g.][]{Bottke_etal_2005a}.
We  fitted an incremental power-law to the two different regions and obtained an index of $-1.2$ for sizes $ \Dc < 110 $~km (solid line, Fig.\,\ref{F1}), and $-4.5$ for $ \Dc > 110 $~km.
The  cumulative distribution functions is then given by
\be
P_{cum.} =  \left\{ 
  \begin{array}{l c r}
   \left(\frac{D_0}{\Da}\right)^{1.2} & \mathrm{if} & D_0 \le \Da \le \Dc \ ,
                    \crm
   \left(\frac{D_1}{\Da}\right)^{4.5} & \mathrm{if} & \Dc < \Da \le 200\mathrm{~km} \ ,
  \end{array} 
 \right.
\llabel{111025c}
\ee
with 
\be
D_0 = \left(\frac{40}{N}\right)^\frac{1}{1.2} \times \mathrm{30~km} 
\quad \mathrm{and} \quad 
D_1 = \left(\frac{D_0}{\Dc}\right)^\frac{1.2}{4.5} \times \Dc \ ,
\llabel{111025d}
\ee
where $ \Dc = 110$~km, and  30~km is the approximate size of a body needed to open a 300~km diameter crater on Mercury (Eq.\,\ref{120304b}).
Thus,
\be
\Da =  \left\{ 
  \begin{array}{l c r}
   D_1 \, x^{-\frac{1}{4.5}} & \mathrm{if} & x \in [0,x_0[ \ ,
                    \crm
   D_0 \, x^{-\frac{1}{1.2}} & \mathrm{if} & x \in [x_0,1] \ ,
  \end{array} 
 \right.
\llabel{111025e}
\ee
where $ x_0 = (D_0 / D_1)^\frac{5.4}{3.3} $.

In our simulations we set $N = 100$, %during the period $[-4.0, -3.8]$~Gyr, the end of the late heavy bombardement \citep{Kring_Cohen_2002}.
%since the Caloris basin, one of the youngest large impact craters, was formed around $-3.7$~Gyr \citep{LeFeuvre_Wieczorek_2011}.
which gives $\log (x N) = \log x + 2$,  $ D_0 \approx 14 $~km, $ D_1 \approx 64$~km, and $ x_0 \approx 8.4
\times 10^{-2}$.
For both ABD and IMD, if $\Da > 200 $~km we repeat the interaction given by expressions (\ref{120408a}) and (\ref{111025e}), respectively.

\begin{table*}
\begin{center}  
\caption{Capture probabilities in several spin-orbit resonances (in percentage).  \llabel{Tab3}} 
\begin{tabular}{c | c c c c | c c | c c} 
\tableline
\tableline
\multirow{2}{*}{$p$} & \multicolumn{4}{c|}{previous studies} &  \multicolumn{2}{c|}{IMD} &  \multicolumn{2}{c}{ABD} \\
 & GP66 & CL04 & CL09 & Wi12 & prograde & retrograde & prograde & retrograde \\
\tableline
  4/1 &   $-$  &   $-$ &   $-$ &    $-$ &   $-$ &  $-$ &   $-$ &  $-$ \\
  7/2 &   0.1  &   $-$ &   4.7 &    $-$ &   $-$ & $-$   &   $-$ & $-$ \\
  3/1 &   0.4  &   $-$ & 11.6 &    $-$ &   $-$&  0.1    &  $-$ & 0.2  \\  
  5/2 &   1.4  &   $-$ & 22.1 &    0.1 &   0.3 &   $-$  &   0.2 &  0.2 \\
  2/1 &   1.7  &   3.6 & 31.6 &    0.4 &   3.7 &   2.5   &   2.9 &   2.7 \\
  3/2 &   7.2  & 55.4 & 25.9 &   2.4  & 49.4 & 52.2  & 50.6 & 50.1  \\ 
  1/1 &   $-$  &   2.2 &   3.9 & 68.2  & 26.9 &  25.9 & 24.2 &  25.1  \\ 
  1/2 &   $-$  &   $-$ &   $-$ & 28.9 &   0.1 &   $-$   &   $-$ &   $-$  \\
none& 89.2 & 38.3 &   0.2 &   $-$ & 19.6  &   19.3 & 22.1  &  21.7 \\ 
\tableline
% 14:24, 28-03-2008
\end{tabular}
\tablecomments{``GP66'' \citep{Goldreich_Peale_1966}: spin evolution only with tides (Eqs.\,\ref{eq1}, \ref{eq2}), constant eccentricity and starting with prograde rotation; ``CL04'' \citep{Correia_Laskar_2004}: same as ``GP66'', but with planetary perturbations; ``CL09'' \citep{Correia_Laskar_2009}: same as ``CL04'', with the addition of CMF (Eq.\,\ref{eqN1}); ``Wi12'' \citep{Wieczorek_etal_2012}: same as ``CL09'', but starting with retrograde rotation.}
\end{center}
\end{table*}

\section{Spin dynamics model}
%\section{Conservative motion}

Tidal dissipation and core-mantle friction drive the obliquity of Mercury close to $0^\circ$ or $180^\circ$ \citep{Correia_Laskar_2010}. 
For zero degree obliquity, 
and in absence of dissipation, the averaged equation for the rotational motion near the $ p $ resonance
(where $ p $ is a half-integer) writes \citep{Goldreich_Peale_1966, Correia_2006}:
\be
\dot \omega = - \frac{3}{2} n^2 \frac{B-A}{C}  H (p, e) \sin 2 (\theta - p M) \ , \llabel{eq1}
\ee
where $ M $ is the mean anomaly,  $ \omega = \dot \theta $,  $H (p, e)$ are Hansen coefficients, and
$A < B < C $ are the moments of inertia.
%The same equation is valid for $180^\circ$ obliquity using negative rotations.
%The libration width of the rotation rate in resonance is \citep{Goldreich_Peale_1966}
%\be
%\Delta \omega = n \sqrt{3 H (p, e) \frac{B-A}{C}} \ . \llabel{120305a}
%\ee

For tidal dissipation we adopt a linear viscous model \citep[e.g.][]{Mignard_1979} %,Goldreich_Peale_1966,Correia_2009}:
in agreement with a Maxwell rheology for slow rotations.
Its contribution to the rotation rate is given by \citep[e.g.][]{Correia_2009}:
\be
\dot \omega = - K \left[ \Omega (e) \omega - N (e) n \right]\ , 
\llabel{eq2}
\ee
with   
\be 
\Omega (e) = \frac{1 + 3 e^2 + 3 e^4 / 8}{(1 - e^2)^{9/2}} \ , \llabel{120305b}
\ee
\be
N (e) = \frac{1 + 15 e^2 / 2 + 45 e^4 / 8+ 5 e^6 / 16}{(1 - e^2)^{6}} \ ,  \llabel{120305c}
\ee
and
\be
K = 3 n \frac{k_2}{\xi \, Q}  \left(\frac{R}{a}\right)^3 \left(\frac{m_\odot}{m}\right) \ ,
\llabel{eq3}
\ee
where $m_\odot$ is the solar mass, $ k_2 $ is the second Love number, $Q$ the quality factor, and $ \xi = C / mR^2 $.

%The equilibrium is achieved when $ \dot \omega = 0 $, that is, for a constant eccentricity $e$, when $ \omega / n =  N (e) / \Omega (e) $. 
%In a circular orbit ($ e = 0 $) this equilibrium coincides with synchronization ($\omega=n$), while the equilibrium rotation rate $ x = 3/2 $ is achieved for $ e_{3/2} = 0.284927 $.
%For this eccentricity, rotation rates smaller than $ 3 n / 2 $ will increase, while higher rotation rates will decrease.

The Mariner~10 flyby of Mercury and subsequent observations made with Earth-based radar provided strong evidence of a conducting existent fluid core \citep{Ness_1978, Margot_etal_2007}.
%The two parts then tend to rotate at different rates, but this tendency is more or less counteracted by friction torques (viscous and electromagnetic) at their interface.
The resulting core-mantle friction (CMF) may be expressed by an effective
torque \citep{Goldreich_Peale_1967,Correia_Laskar_2010}:
\be
\dot \omega = - c_c \kappa (\omega-\omega_c) \quad \mathrm{and} \quad \dot \omega_c = c_m \kappa
(\omega-\omega_c) \ , \llabel{eqN1} 
\ee
where $ \kappa $ is an effective coupling parameter, $ \omega_c $ the core's rotation rate, and  $ c_c = C_c / C = 1 - c_m = 0.45 $ \citep{Margot_etal_2007}. 
We also have $ \kappa = 2.62  \sqrt{\nu \omega} / (c_m R_c) $ \citep{Mathews_Guo_2005}, where $R_c $ is the core radius and $ \nu $ is the kinematic effective viscosity of the core.
%An analytical approximation for the capture probability into resonance $p$, assuming an even distribution in energy is given by\cite{CoLa09,GoPe67}
%\be
% P_\mathrm{cap}^\pm = 2 \, (1+\chi) \left[ 1 \pm c_m \frac{\pi}{2}  \frac{p - x_l(e)}{\Delta_p x} \right]^{-1}  \ , \llabel{eq4} 
%\ee
%where $ \chi = c_c \kappa / [ K \Omega(e) ] $ and $ P_p^+ $ (resp. $ P_p^- $) is the probability of capture when the spin is decreasing (resp. increasing).

%We adopt for Mercury's geophysical parameters \citep{Anderson_etal_1987, Poirier_1988, Goldreich_Soter_1966, Spohn_etal_2001}:
%$ (B-A)/C = 1.2 \times 10^{-4} $, $ R_c / R = 0.77 $,
%$\xi=0.3333$, $k_2 = 0.4 $, $Q = 50$ and $ \nu = 10^{-6} \mathrm{m}^2
%\mathrm{s}^{-1} $ (which yields %for the 3/2 resonance
%$ K \approx 8.4 \times 10^{-7} \, \mathrm{yr}^{-1} $ and $ \kappa \approx 5 \times 10^{-5}
%\, \mathrm{yr}^{-1} $).

When considering the perturbations of the other planets, the eccentricity of Mercury undergoes strong chaotic variations in time \citep{Laskar_1994, Laskar_2008, Correia_Laskar_2004, Correia_Laskar_2009}. 
These variations are modeled using the averaging of the equations for the motion of the Solar System, that have been compared to numerical integrations, with very good agreement \citep{Laskar_etal_2004M,Laskar_etal_2004E}.
The mean value of the eccentricity is $0.198 $, slightly lower than the present value, but we also observe a wide range for the eccentricity variations, from nearly zero to more than 0.45. 
Even if some of these episodes do not last for a long time, they will allow additional
capture into and escape from spin-orbit resonances (Table \ref{Tab2} and \ref{Tab3}).
%Indeed, if $ K [ \Omega (e) p - N (e) ] / H(p,e) > 3 n K (B-A) / (2 C) $ the equilibrium in the $p$ spin-orbit resonance can no longer be sustained and the resonance is destabilized\cite{CoLa04,CoLa09}.
%The synchronous resonance is always stable, but critical eccentricities for other resonances are $ e_c = 1.8 \times 10^{-4}$ for the 1/2 resonance, $ e_c = 2.6 \times 10^{-5} $ for the 3/2 and $ e_c = 4.6 \times 10^{-3} $ for the 2/1.

\section{Numerical simulations}

Assuming an initial rotation period of Mercury of 8~h, we estimated the time needed to despin the planet to the slow rotations to about 300 million years. 
This motivates that our starting time is $-4$~Gyr, although this value is not  critical.
Due to the chaotic behavior of the eccentricity \citep{Laskar_1989,Laskar_1990}, we have performed a statistical study of the past evolutions of Mercury's orbit, with the integration of 1000 orbits in the past, starting with very close initial conditions, within the uncertainty of the present ones. 
For each of these 1000 orbital solutions, we have integrated numerically the rotation motion of Mercury, taking into account the  resonant terms (Eq.\ref{eq1}), for $p=k/2; k=-14,\dots,14$, the tidal dissipation (Eq.\ref{eq2}), the CMF (Eq.\ref{eqN1}), and the planetary perturbations.
We adopted for all model parameters the same values as in \citet{Correia_Laskar_2009}.
%, starting with a prograde (sim #1) and retrograde (sim #2) rotation period of
%$\pm 10$~days. %($\omega_0 / n \approx -8.8$).  
The effect from the collisions (Sect.\,\ref{sectcoll}) is only considered between $-4.0$ and $-3.8$~Gyr in the past. %, i.e., it ceases after the late heavy bombardment \citep{Kring_Cohen_2002}.

\begin{figure*}%[htb]	% h-here, t-top, b-bottom
\centering
\includegraphics[width=18.cm]{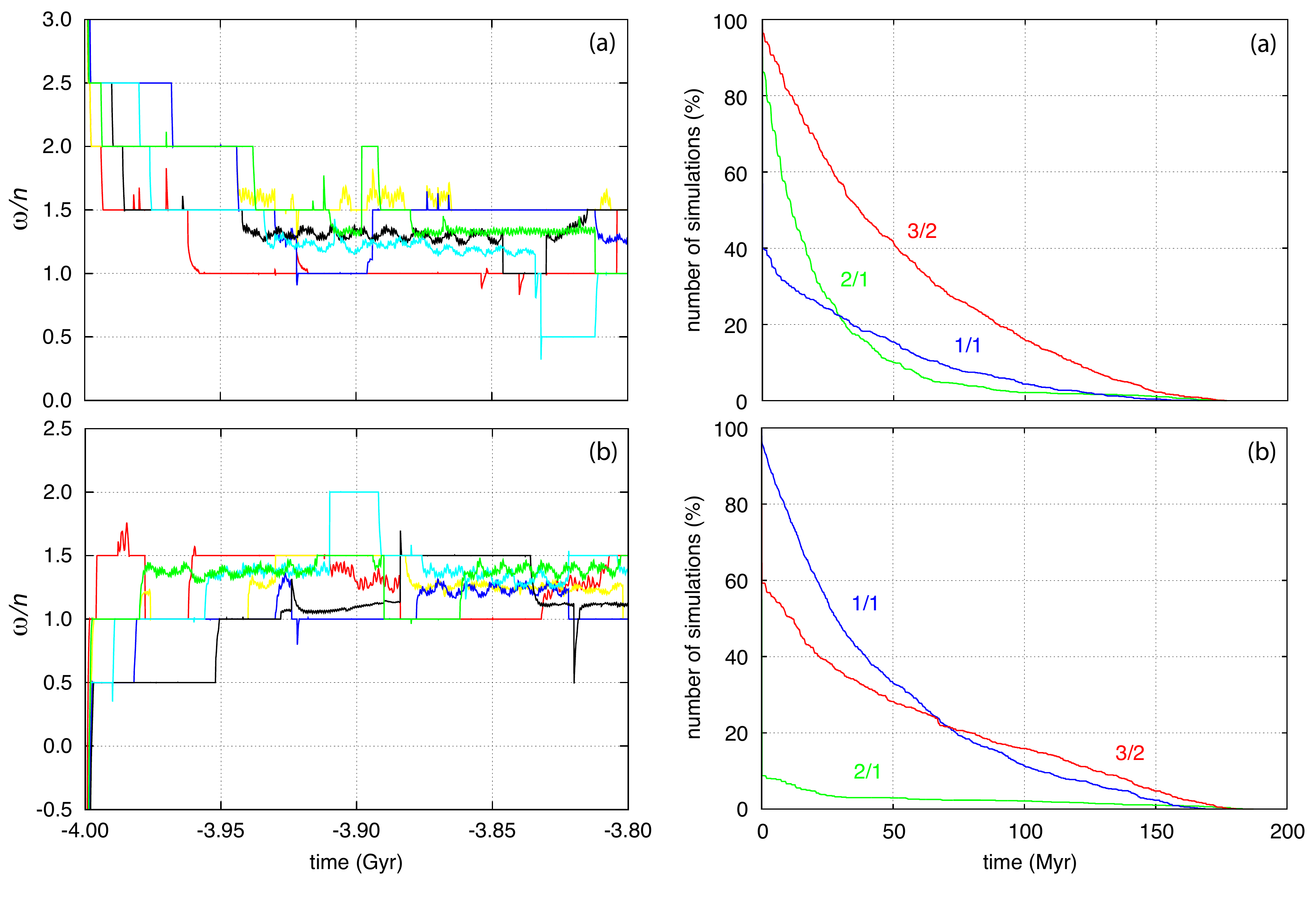} %	** if .eps don't need extension
\caption{{\bf Left:} Some examples of Mercury's rotation rate evolution during the collisional period. {\bf Right:} Cumulative time spent in each spin-orbit resonance (IMD). {\bf (a)} For initial prograde rotation. {\bf (b)} For initial retrograde rotation. \llabel{F2}}
\end{figure*}

In a first experiment, we use the IMD for collisions, and the initial rotation period of Mercury is set at 10~day (prograde rotation).
As in previous studies, the spin encounters higher order resonances first, and capture often occurs (Fig.\,\ref{F2}a).
%Sometimes the moment imparted by an impact is enough strong to disrupt the resonant equilibrium.
Small impacts cause the planet to librate in longitude about the resonance center, with the amplitude damping with time, but for sufficiently large impacts, escape from all resonances may occur (Tab.\,\ref{Tab2}).  
Half of these breaking events increase the rotation rate and the spin can be recaptured in the same resonance again.
For the other half, the spin progressively reaches its equilibrium position $ \omega / n \sim N(e) / \Omega(e) $ (Eq.\,\ref{eq2}), which corresponds to $ \omega / n \sim 1.25 $ for the average eccentricity of Mercury.
Therefore, the final impacts more likely drive the spin into the the closest resonances, i.e., the 1/1 or the 3/2 resonance.
As a consequence, contrarily to previous studies (Table\,\ref{Tab3}), it becomes possible to reach the synchronous resonance starting with prograde rotation (Fig.\,\ref{F2}a).

After the end of the collisional period (at $-3.8$~Gyr), about half of the simulations follow the equilibrium rotation rate (i.e., they are not trapped).
However, the subsequent evolution due to the chaotic behavior of the eccentricity still leads many of these solutions to capture \citep{Correia_Laskar_2004}.
At the end of the simulations, the 3/2 resonance presently observed becomes the most probable outcome (half of the simulations), followed by one quarter in the synchronous resonance (Table\,\ref{Tab3}).
Notice also that about 40\% of the final captures in the present equilibrium experienced some time in the synchronous resonance (Fig.\,\ref{F2}a), a scenario compatible with the observed crater distribution  \citep{Wieczorek_etal_2012}.

Keeping the IMD for collisions, we then repeated all the 1000 simulations, but starting the rotation period of Mercury at $-10$~day (retrograde rotation).
Now, lower order resonances are encountered first.
We did not observe any capture in negative resonances, and the 1/2 resonance is easily destabilized by collisions (Table\,\ref{Tab2}).
However, there is a strong chance of capture in the 1/1 resonance, and about 95\% of the simulations spend some time there (Fig.\,\ref{F2}b).
Once in the synchronous resonance, half of the impacts tend to increase the rotation rate to higher values, so higher order resonances are also achievable for initial retrograde planets.
Again, after the end of the collisions (at $-3.8$~Gyr), about half of the simulations are following the equilibrium rotation rate, so the subsequent evolution is very similar to initial prograde rotation.
As a consequence, the final distribution of capture probabilities are identical in the two scenarios (Table\,\ref{Tab3}).

Finally, we repeated the 1000 simulations for both initial prograde and retrograde rotation, but using the ABD for collisions. 
Here, spin-orbit resonances are destabilized more often, since the ABD increases the size of the bodies that collide with Mercury for $\Da < \Dc$.
However, the final statistics are very close (Table\,\ref{Tab3}), 
showing that the collisional model is not a critical parameter for the spin evolution.

\section{Consclusion}

The observed impact basins on Mercury's surface allows us to reconstruct the consequences of the late heavy bombardment on this planet.
%The fossilized size distribution of impacts on Mercury's surface allows us to reconstruct the consequences of the late heavy bombardment on this planet.
%First, we show that this distribution matches quite well the theoretical initial size distribution of the main-belt. 
Using a collisional model based on two different size distributions, we simulate the effect of the impacts on the spin evolution of the planet.
We observe that the present 3/2 resonant state becomes the most probable outcome for the rotation (about 50\% of chances).
This value is twice larger than when collisions are not considered \citep{Correia_Laskar_2009}, but very similar to the case where CMF is also neglected \citep{Correia_Laskar_2004}.
The final distribution in each spin-orbit resonance is then more sensitive to the orbital solution statistics, rather than to the tidal, CMF, or collisional models that we use.
Indeed, when collisions are taken into account, all initial captures in spin-orbit resonances are destabilized, and the final evolution of the eccentricity preferably sets the rotation of the planet between the present state and the synchronous resonance.
Therefore, we do not expect that the capture probability values given in this paper will change much if other dissipative models are adopted in the future \citep[e.g.][]{Makarov_2012}.
In addition, the collisional model presented here is also able to reproduce a temporary capture of the planet in the synchronous resonance \citep{Wieczorek_etal_2012}, without needing any particular assumption on the initial orientation of the spin. %rotation rate.
%The fact that we are able to explain many different features observed on Mercury, puts strong constraints on the initial distribution of asteroids in the main-belt, in particular it validates \citet{Bottke_etal_2005a} model.

\acknowledgments

We acknowledge support from PNP-CNRS, CS Paris Observatory, CNRS-PICS05998,
and FCT-Portugal (PTDC/CTE-AST/098528/2008, SFRH/BSAB/1148/2011, PEst-C/CTM/LA0025/2011). 

%\bibliographystyle{apj}
%\bibliography{correia}

\end{document}